\documentclass[reprint,amsmath,amssymb,aps,prl]{revtex4-1}
\usepackage[dvips]{graphicx}
\usepackage{dcolumn}
\usepackage{bm}
\usepackage{hyperref}
\usepackage{color}
\usepackage{cancel}
\usepackage{bmpsize}
\usepackage{amsmath}
\usepackage{amssymb}

\newcommand{\bra}[1]{\left\langle #1\right|}
\newcommand{\ket}[1]{\left| #1\right\rangle}
\newcommand{\braket}[2]{\left\langle
#1\vphantom{#2}\right|\left.#2\vphantom{#1}\right\rangle}
\newcommand{\ketbra}[2]{\left| #1\right\rangle\!\left\langle#2\right|}

\newcommand{\be}[0]{\begin{equation}}
\newcommand{\ee}[0]{\end{equation}}

\newcommand{\lra}\simeq
\newcommand{\eeqref}[1]{Eq.~(\ref{#1})}

\begin{document}

\title{Loss-tolerant quantum enhanced metrology and state engineering \\ via the reverse Hong-Ou-Mandel effect.}

\author{Alexander E. Ulanov$^{1,2}$, Ilya A. Fedorov$^{1,3}$, Demid Sychev$^{1}$, Philippe Grangier$^{4}$ and Alexander I. Lvovsky$^{1,3,5}$}

\affiliation{$^1$Russian Quantum Center, 100 Novaya St., Skolkovo, Moscow 143025, Russia}
\affiliation{$^2$Moscow Institute of Physics and Technology, 141700 Dolgoprudny, Russia}
\affiliation{$^3$P. N. Lebedev Physics Institute, Leninskiy prospect 53, Moscow 119991, Russia}
\affiliation{$^4$Laboratoire Charles Fabry, Institut d'Optique Graduate School, CNRS, Universit\'e Paris-Saclay, 91127 Palaiseau, France}
\affiliation{$^5$Institute for Quantum Science and Technology, University of Calgary, Calgary AB T2N 1N4, Canada}

\email{LVOV@ucalgary.ca}
\date{\today}
%
\begin{abstract}

Preparing highly entangled quantum states between remote parties is a major challenge for quantum communications \cite{Our09,Dakic2012,Lopaeva2013,Ulanov15,Simon2003,Monroe,Weinf,Hanson}.
Particularly promising in this context are the N00N states, which are entangled $N$-photon wavepackets delocalized between two
different locations, providing measurement sensitivity limited only by the uncertainty principle \cite{Lee2001,Nagata2007,Edamatsu2002,Chen2015,Dowling2009,Kacprowicz2010,Rubin2007}.
However, these states are notoriously vulnerable to losses, making it difficult both to share them between remote locations, and to recombine them to exploit interference effects. Here we address this challenge by utilizing
the reverse version of the Hong-Ou-Mandel effect \cite{HOM87} to prepare a high-fidelity two-photon N00N state shared between two parties connected by a lossy optical channel. Furthermore, we demonstrate that the enhanced phase sensitivity can be directly exploited in the two distant  locations, and  we  remotely prepare superpositions of coherent states, known as ``Schr\"odinger's cat states" \cite{Ourjoumtsev2007,AndersenRev}.

\end{abstract}

\maketitle
\vspace{10 mm}

In the current race towards the practical implementation of quantum techniques for  information processing and communications, a strong  trend is to design
loss-tolerant quantum protocols, such as the preparation of non-local superpositions of quasi-classical light states \cite{Our09},
discord-assisted remote state preparation \cite{Dakic2012},  quantum illumination \cite{Lopaeva2013}, undoing the effect of losses on continuous-variable entanglement \cite{Ulanov15}, and the preparation of single qubit entangled states over a long distance \cite{Simon2003,Monroe,Weinf,Hanson}.

In this article we will be interested in N00N states $\ket{N::0}= (\ket{N,0}+\ket{0,N})/\sqrt{2}$, which are useful  in linear-optical quantum computation \cite{Dowling2005, Lim2005}, quantum-optical state engineering \cite{AndersenRev,Etesse2015}, and the preparation of photon-number path entanglement \cite{Lee2001,Nagata2007}.  But the most important potential application of these states is as a resource for quantum-enhanced metrology \cite{Edamatsu2002, Walther04, Mitchell04, Nagata2007, Dowling2009, Chen2015}. Interference measurements with N00N states exhibit super-resolving properties: the number of  fringes per wavelength equals $N$, in contrast to a single fringe in the case of coherent states. This property can be exploited for precise measurement of diverse physical quantities.
Widespread application of N00N states for metrology is however precluded by their extreme sensitivity to losses. When exposed to even moderate losses, the degree of  entanglement, and hence the super-resolution potential of the N00N states dramatically degrades to an extent that eliminates any advantage  \cite{Kacprowicz2010, Rubin2007}.

In the present work, we address this challenge by developing a technique to losslessly produce N00N states between parties that are separated by a lossy quantum channel. In addition, using N00N states usually requires to bring back together the two entangled parts, introducing more  propagation losses. But we will show that this second step is not required, and that super-resolution for the optical phase  can be obtained remotely, by using homodyne detection \cite{AndersenRev}.

For N00N state production,  we exploit some peculiar properties of the Hong-Ou-Mandel (HOM) effect \cite{HOM87}, a well-known quantum interference phenomenon in which two indistinguishable photons that are overlapped on a symmetric beam splitter (BS) always emerge in the same output mode,  preparing  the N00N state
\begin{equation}
\begin{aligned}
\label{eq1}
& \ket{11} \rightarrow \ket{2::0}=  \dfrac{\ket{2,0}+\ket{0,2}}{\sqrt{2}}
\end{aligned}
\end{equation}
in the beam splitter output.
Our experiment relies upon the \emph{reverse HOM effect}, in which the measurements in the two output modes of the BS project them onto single-photon states. Because of the time-reversible nature of quantum mechanics, such projection is equivalent to projecting the state of the input onto the two-photon N00N state \eqref{eq1}. If each of the beam splitter inputs is, in turn, entangled with other modes, these modes become entangled with each other thanks to entanglement swapping \cite{Pan1998}. Whereas the original HOM setting creates a two-photon N00N state, straightforward extensions will produce
$\ket{N::0}$ states (see Supplementary Material \cite{SuppInfo}).

\begin{figure}[h]
	\includegraphics[width=0.95 \columnwidth]{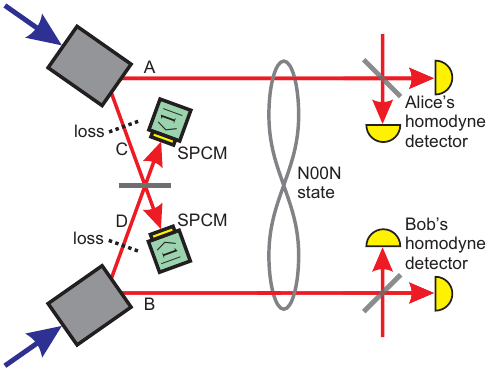}
	\caption{Conceptual scheme of the experiment. The two-photon N00N state of modes A and B is produced in the event of a coincidence click of the SPCMs.}
	\label{f1}
\end{figure}

Specifically, consider a setup in which two pairs of modes (A,C) and (B,D) are prepared in two-mode squeezed states
\begin{equation}
\label{eq2}
 \ket{\Psi}_{AC/BD}\propto\ket{0,0}+\gamma\ket{1,1}+\gamma^{2}\ket{2,2}+...
\end{equation}
by means of non-degenerate parametric down-conversion.
Modes C and D are then mixed on a symmetric BS, outputs of which are subjected to measurement in the photon number basis via single-photon counting modules (SPCMs), as shown in Fig.~\ref{f1}.
\\

In the weak-squeezing limit $|\gamma|^2\ll\eta_{\rm SPCM}$, where $\eta_{\rm SPCM}$ is the SPCM's efficiency, every SPCM click is likely to be caused by no more than one photon. Then, a coincidence click in both SPCMs correspond to projection on state $\ket{1,1}_{CD}$. Due to the unitary nature of the BS operation, this event assures that modes C and D were initially prepared in the N00N state (\ref{eq1}). This corresponds to two pairs of photons having been produced in either of two crystals; therefore, the remaining modes A and B now also share a two-photon N00N state.

\begin{figure} [b]
	\includegraphics[width=3in]{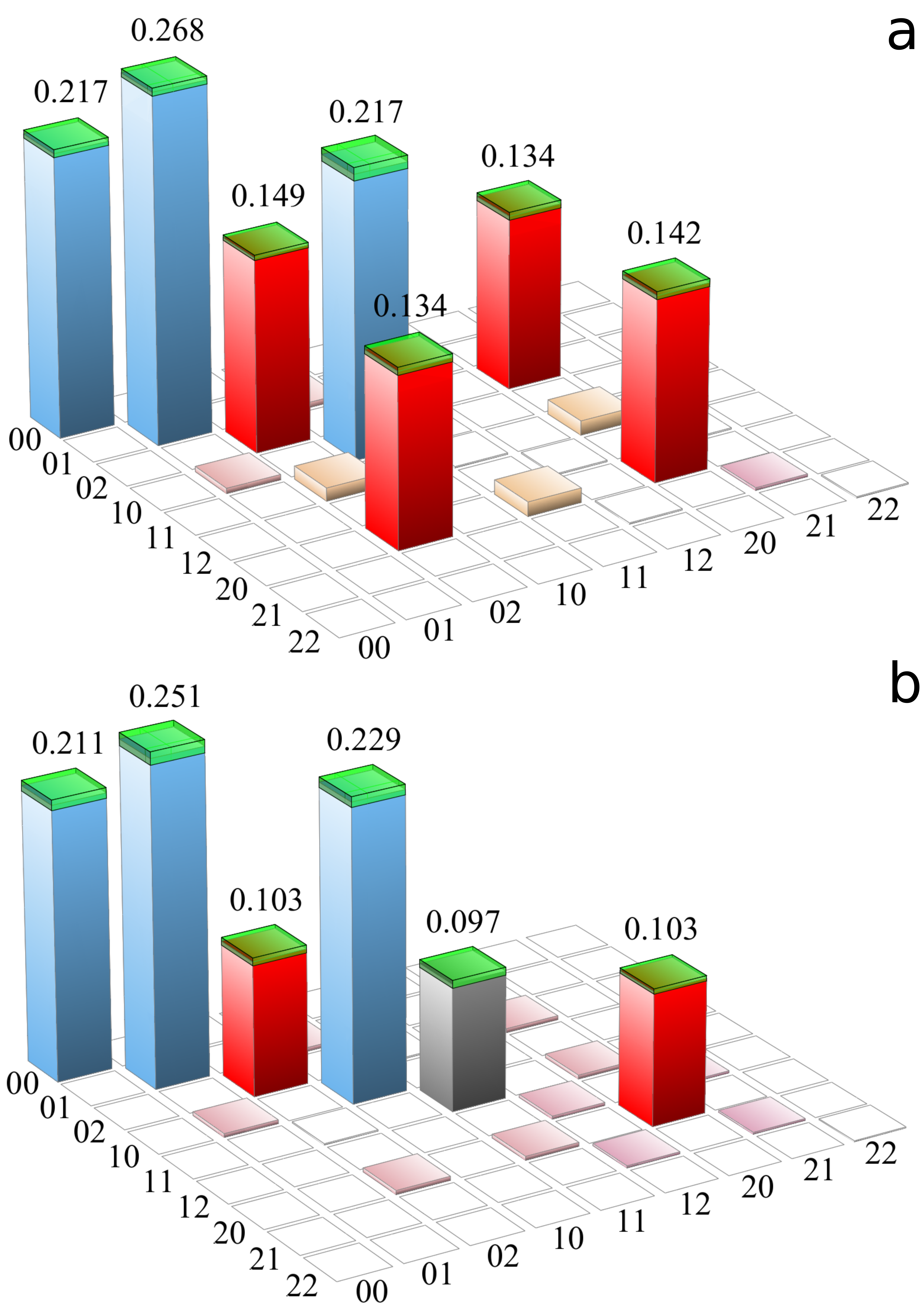}
	\caption{(a) Density matrix of the state of modes A and B, with a 10-dB total loss in modes C and D before the BS, reconstructed via the maximum likelihood algorithm \cite{Lvo2004,Lvo2007} in the Fock basis. The numbers along the horizontal axes are the indices of the bra and ket elements of that matrix. The components shown in red correspond to the ideal N00N state, others appear due to losses. The $\ket{1,1}\bra{1,1}$ component amounts to $0.01$. (b) Density matrix in the case of a $5$-ps mismatch between modes C and D,  eliminating  HOM interference. The emerged $\ket{1,1}\bra{1,1}$ component is shown in gray. Green hats show the statistical error of the reconstruction.}
	\label{DM}
\end{figure}

A remarkable feature of this scheme its robustness to losses in channels C and D. Indeed, such losses are equivalent to corresponding reduction of the SPCM's quantum efficiencies \cite{Berry2010}. If the down-conversion amplitude $\gamma$ is sufficiently small, the leading term in the state of channels $A$ and $B$ conditioned on the SPCMs' coincidence click is still the two-photon N00N state. One can therefore eliminate the effect of loss, however high it may be, on the purity of the N00N state.

Conditioned on coincidence clicks of the two SPCMs (see Methods), we characterize the state of modes A and B by means of homodyne tomography \cite{Kumar2012,Lvovsky2009}. The measured states in both cases are very close to the ideal state (\ref{eq1}) that has suffered a $1-\eta=45\%$ loss due to imperfect detection efficiency \cite{Lvovsky2001, Huisman2009} [Fig.~\ref{DM}(a)].

To illustrate the reverse HOM effect, we measured the behavior of the state in modes A and B as a function of increasing temporal mismatch between modes C and D. As evidenced by Fig.~3, the fraction of the biphoton component $\ket{1,1}$ in that state exhibits a dip that is characteristic as a signature of the HOM effect. The tomographic state reconstruction  for the case of complete mismatch is shown in Fig.~\ref{DM}(b). In addition to a macroscopic biphoton component, this state exhibits no off-diagonal terms because no coherence between modes A and B can emerge in the absence of interference.

\begin{figure} [t]
	\includegraphics[width=0.8\columnwidth]{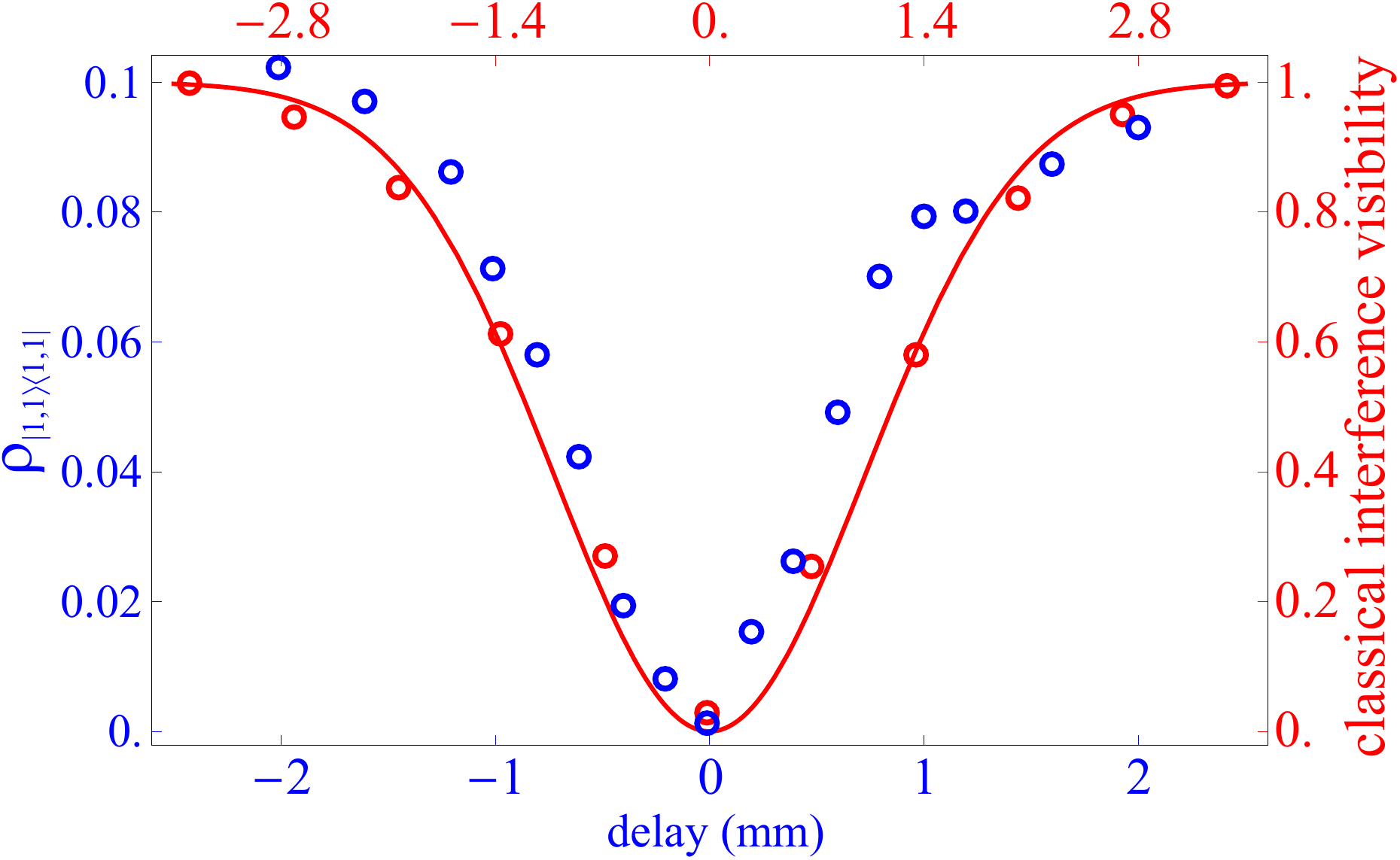}
	\caption{Quantum and classical interference as a function of the delay between modes C and D. The blue points represent the weight of the $\ket{1,1}\bra{1,1}$ component in the reconstructed density matrix (obtained from homodyne tomography, not from photon counting). The red points and red curve correspond to the visibility of the classical interference between the master laser pulses that underwent the same spectral filtering as quantum modes C and D. The horizontal axes for the classical and quantum cases are scaled by a factor of $\sqrt 2$ with respect to each other to match the theoretically expected widths of the interference patterns.}
	\label{HDip}
\end{figure}

The enhanced phase sensitivity is manifested by the mean values of observables $X_A X_B$ and $X_A^2 X_B^2$, where $X_A = Q_A\, \mathrm{cos} \, \theta_A + P_A\, \mathrm{sin} \, \theta_A$ and $X_B = Q_B\, \mathrm{cos} \, \theta_B + P_B\, \mathrm{sin} \, \theta_B$ are quadrature operators of modes A and B. For the one-photon $\ket{1::0}= (\ket{1,0}+\ket{0,1})/\sqrt{2}$ and two-photon N00N states one has, respectively,
\begin{equation}
\begin{aligned}
\label{eq41}
 \bra{1::0}X_A X_B\ket{1::0} &= -\frac{\eta}{2} \mathrm{sin}\, \triangle \theta, \\
 \bra{1::0}X_A^2 X_B^2\ket{1::0} &= \frac{ 1 + 2 \eta}{4} \\
\end{aligned}
\end{equation}
and
\begin{equation}
\begin{aligned}
\label{eq42}
\bra{2::0}X_A X_B\ket{2::0} &= 0, \\
 \bra{2::0}X_A^2 X_B^2\ket{2::0} &= \frac{1}{4} + \eta + \frac{\eta^2}{2} \, \mathrm{cos}\,(2 \triangle \theta)
\end{aligned}
\end{equation}
where $\quad \triangle \theta = \theta_A-\theta_B$.

We see that in order to compare between the one- and two-photon N00N states using quadrature measurements, we need to use different observables. This notwithstanding, the phase dependence of the appropriate observable in each state is as expected: with period $2\pi/N$ for each $N$-photon N00N state.

\begin{figure}[t]
	\includegraphics[width=0.8\columnwidth]{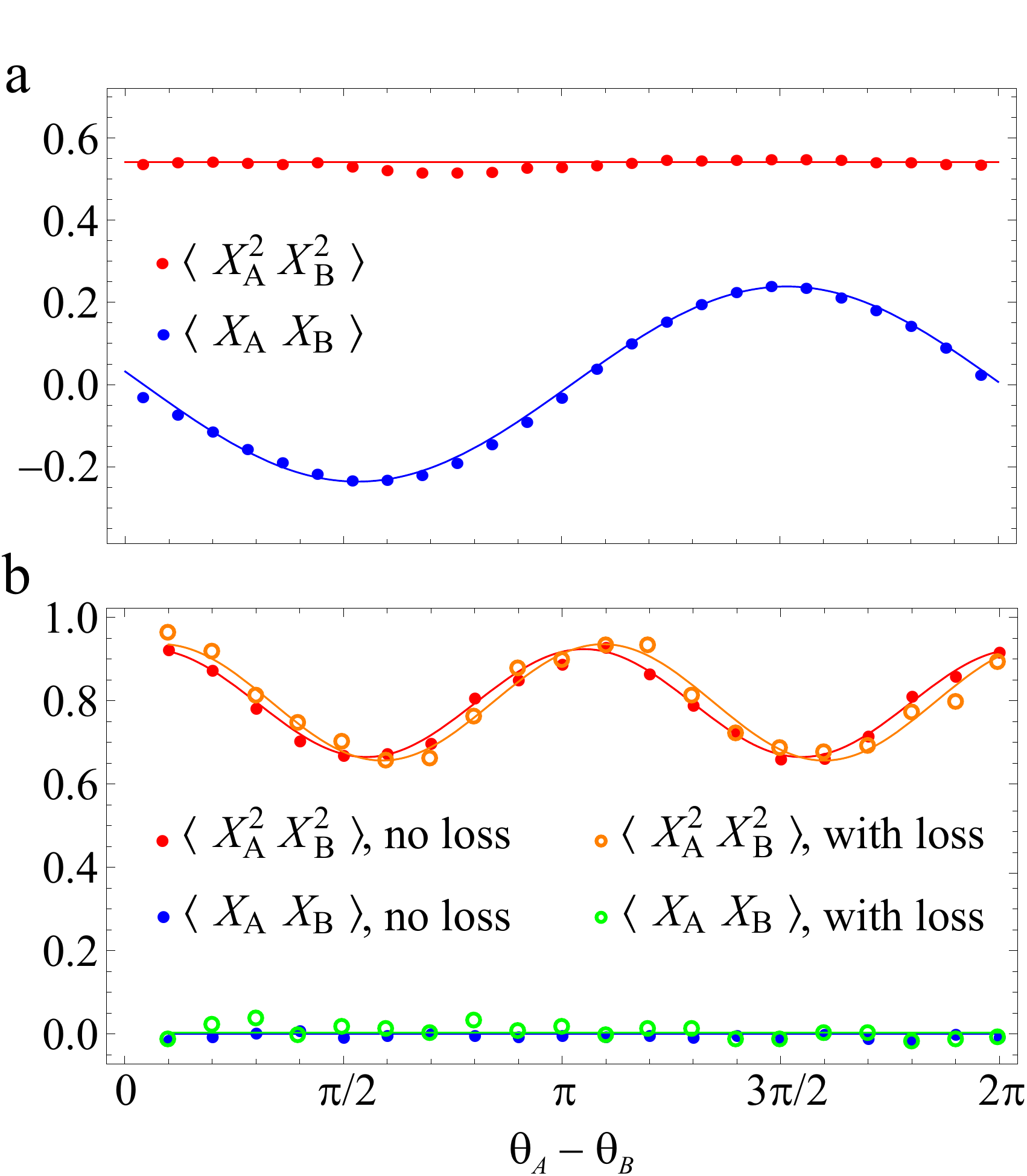}
	\caption{Dependence of the mean  and mean square product of Alice's and Bob's quadratures on the difference $\theta_A-\theta_B$ of Alice's and Bob's phases for states $\ket{1001}$ (a) and $\ket{2002}$ (b). Enhanced phase sensitivity is evident for the two-photon N00N state. Solid lines are theoretical predictions.}
	\label{CORR}
\end{figure}

An experimental check of this behavior is demonstrated in Fig.~\ref{CORR}. In agreement with the results of state reconstruction, the second-order interference of the $\ket{2002}$ state, generated in the presence of the loss, exhibits the same visibility as without loss.

\begin{figure*}[t]
	\includegraphics[width=0.975 \textwidth]{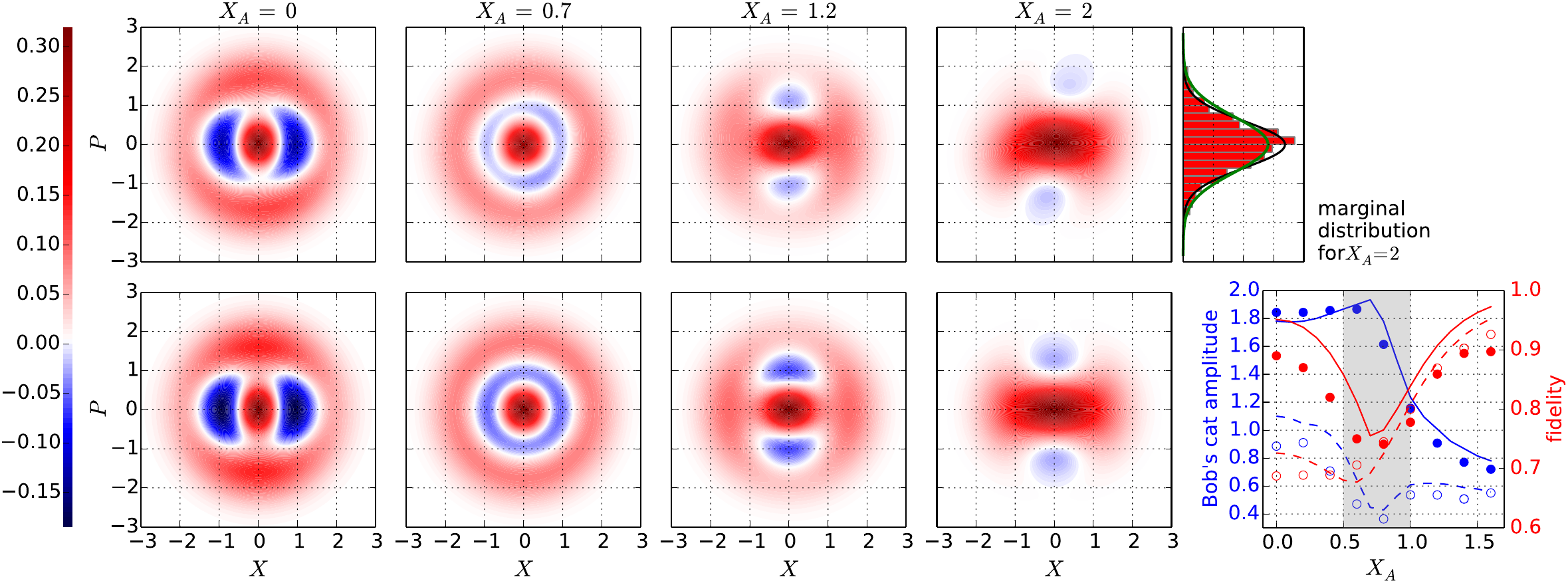}
	\caption{Top row: Bob's Wigner functions after conditioning on Alice's quadrature measurement reconstructed from experimental data with 55\% efficiency correction. Bottom row: theoretical calculation. Fidelities between experiment and theory are $0.98$, $0.98$, $0.97$, $0.99$ from left to right. The top-right panel shows the marginal distribution of Bob's position quadrature in the case of $X_A = 2$, which exhibits squeezing. The black line fits the data, the green line corresponds to the vacuum. The bottom rightmost panel shows the amplitude of the CSS that best approximates the state of Bob's mode versus Alice's quadrature, as well as the corresponding fidelity. Lines represent theoretical calculations. Filled (empty) markers and solid (dashed) lines stand for the data with (without)  the efficiency correction. The approximation of Bob's state with a CSS state near $X=0.7$ (second column) is unreliable because of the unproportionally large two-photon fraction and low phase-dependence; this instability causes opposite behavior of the blue lines in the shaded region.}
	\label{WIGS}
\end{figure*}

Being path-entangled, the N00N state can also be used for single-mode quantum-state engineering. Associating modes A and B with fictitious observers Alice and Bob, we consider a setting in which Alice, by performing quadrature measurements on her mode, remotely prepares a state in Bob's mode. Neglecting inefficiencies, Alice's quadrature outcome $X$ measured at phase angle $\theta$ brings the Bob's mode to state
\begin{equation}
\begin{aligned}
\label{eq6}
 _A\!\braket{X_\theta \,}{\!2002\,}\!_{AB}
=   _A\!\braket{X_\theta \,}{2}_A\ket{0}_B + _A\!\braket{X_\theta \,}{0}_A\ket{2}_B
\end{aligned}
\end{equation}
where
\begin{equation}
\begin{aligned}
\label{eq5}
\braket{X_\theta \,}{m}_A = e^{im\theta}e^{-X^2}\frac{H_m(X)}{\sqrt{2^{m-1/2} \, m!}},
\end{aligned}
\end{equation}
are Fock state wavefunctions, with $H_m(X)$ being Hermite polynomials. In this way, by postselecting on specific values of Alice's observed quadrature, one can generate arbitrary superpositions of the 0- and 2-photon states, which approximate the even coherent-state superpositions (CSS) $\ket\alpha+\ket{-\alpha}$, sometimes viewed as
a quantum optical implementation of the Schr\"odinger's cat paradox \cite{Ourjoumtsev2007,AndersenRev,Huang2015,Etesse2015}.

The states of Bob's mode for different outcomes of Alice's homodyne measurement are displayed in Fig.~\ref{WIGS}. Projection on $X = 0$ (Fig.~\ref{WIGS}, first column) results in superposition $0.52\ket{0}-0.85\ket{2}$, partially mixed due to losses in Alice's channel. After correcting for Bob's homodyne detection inefficiency, this state has fidelity of $0.88$ with the  even CSS state of amplitude $\alpha=1.84$, squeezed with parameter $z=0.48$. This value compares favorably with state-of-the-art results \cite{Etesse2015, Huang2015}, with the added advantage that our protocol is tolerant to the losses in the optical channel between Alice and Bob.

Effective CSS amplitudes and approximation fidelities for other values of $X$ are shown in Fig.~\ref{WIGS}, right bottom panel.
With increasing $X$, the two-photon fraction initially increases relative to the vacuum because the two-photon state wavefunction $\braket X 2$ decreases faster than the vacuum wavefunction $\braket X 0$. Wavefunction $\braket X 2$ changes sign near value of $X = 0.7$, where ideally a pure two-photon state in Bob's mode should be observed. In practice, due to the losses and finite width of Alice's post-selection window, a phase-insensitive mixture of $0.4\ketbra{0}{0}+0.6\ketbra{2}{2}$ is produced (second column). In this region,  the remotely prepared state approximates the CSS poorly because of the high two-photon component. For higher values of X, this two-photon component in Bob's state reduces again, resulting in increasingly faithful approximation of even CSSs with decreasing amplitudes (third column). For very low amplitudes, this state approximates a weakly-squeezed vacuum state. This is the case for $X = 2$ (fourth column): Bob's quadrature spectrum exhibits squeezing by $0.65\pm0.24\mathrm{dB}$ (without efficiency correction). The corresponding quadrature histogram is shown in the upper right panel of Fig.~\ref{WIGS}.

The protocol developed here addresses the primary challenge in the way of employing nonclassical states of light, particularly the N00N state, for quantum metrology: optical losses. With a conventional fiber channel, it allows establishing nearly ideal N00N entanglement over tens to hundreds of kilometers, no matter how high the loss in the channel connecting the parties is.

Our scheme could be used, for example, for precise phase locking of  local   frequency standards at Alice's and Bob's sites. To realize that, Alice and Bob would employ local oscillators that are referenced to these standards. Then they  would both prepare two-mode squeezed states and perform homodyne measurements on one of the modes while sending the other one to a third party located midway between them for a reverse HOM measurement. When that party obtains a coincidence event, it would inform Alice and Bob via a classical channel, in which case they would record the measured quadrature values. The recorded data would subsequently be used to estimate the phase difference between the local oscillators, eliminating  the contribution of the optical path by closely repeated measurements. The result would then be fed back to one of the frequency standards.

The scheme can be generalized to higher order N00N states, and as shown above it can also be used for ultra-remote preparation of quantum states of light, with possible applications in quantum cryptography  \cite{SuppInfo}. It is therefore a valuable addition to the quantum state engineering  toolbox, for both discrete- and continuous-variable degrees of freedom of optical modes \cite{AndersenRev,Jeong14,Morin14}.

\section{Methods}
We employ a pulsed Ti:Sapphire laser (Coherent Mira 900D) with a wavelength of 780 nm, mean power 1.3 W, repetition rate of 76 MHz and a pulse width of $\sim 1.6$ ps. Most of the laser output is directed into an LBO crystal for frequency doubling. We obtain
up to 300 mW second harmonic; after subsequent spectral cleaning, about 100 mW remain. Then we implement parametric down-conversion in two periodically poled potassium titanyl phosphate crystals in a type II spectrally and spatially degenerate, but polarization non-degenerate configuration. The single photon detection is implemented using SPCMs by Excelitas. The modes entering the SPCMs are spectrally filtered and delivered to the detectors by means of single-mode fibers, which ensures proper preparation of the heralded photon mode \cite{Lvovsky2001, Huisman2009}. Including the filters and fibers, the quantum efficiency of these detectors is estimated as $\eta_{\rm SPCM}\sim 0.15$.

Without losses, the experimental double-coincidence event rate is $\sim$ 100 Hz, which corresponds to a probability of $\sim 10^{-6}$ per pulse and the down-conversion amplitude $\gamma^2\sim0.007$. When we introduce a 10 dB total loss equally distributed between modes C and D, the coincidence rate decreases by a factor of $10$ due to the reduction of the equivalent SPCM efficiency down to $\sim 0.1$. The condition $|\gamma|^2\ll\eta_{\rm SPCM}$ therefore holds even in the presence of losses.

\section{Acknowledgements}
We thank J. Dowling, Y. Kurochkin, A. Pushkina and A. Fedorov for helpful discussions and the Russian Quantum Center for support.

\newpage
\makeatletter
\@addtoreset{equation}{section} 
\makeatother
\makeatletter
\@addtoreset{figure}{section} 
\makeatother
\makeatletter
\renewcommand{\thefigure}{S\@arabic\c@figure}
\makeatother
\section{Supplementary material}

\subsection{Preparation of higher-order N00N states}
In the main text, we have shown that projecting modes C and D from Alice's and Bob's parametric amplifiers onto the two-photon N00N state collapses the counterpart pair of modes, A and B, onto the same state. Here we theoretically discuss the extension of this method to higher photon numbers.

The scheme for projecting modes C and D onto the 4-photon N00N state is shown in Fig.~1(a). Overlapping two modes of that state, which we associate with annihilation operators $\hat c$ and $\hat d$, on a symmetric beam splitter will transform it into
$$\ket{4::0}=\dfrac{\ket{4,0}+\ket{0,4}}{\sqrt{2}}\to \frac{\ket{3,1}+\ket{1,3}}{\sqrt{2}}.$$
Subsequently, we implement photon subtraction in each channel by placing a weakly transmitting beam splitter therein and conditioning on clicks in the transmitted channels. This produces the 2-photon N00N state
$$\hat c\hat d\frac{\ket{3,1}+\ket{1,3}}{\sqrt{2}}=\sqrt{\frac 32}(\ket{2,0}+\ket{0,2}),$$
which can be detected using the reverse Hong-Ou-Mandel effect. Altogether, a coincidence click of all four SPCMs projects means that the 4-photon N00N state has been present at the input.

This scheme works perfectly if the detectors are number-discriminating. Otherwise one of the SPCMs may click in response to more than one photon. Such an event would imply that modes C and D have initially contained five or more photons, and hence have not been in the N00N state. This situation can be avoided by pumping the parametric amplifiers at low powers so that the probability of producing $N+1$ pairs in the two crystals is negligible compared to that of producing $N$ pairs.

\begin{figure}[h]
	\includegraphics[width=\columnwidth]{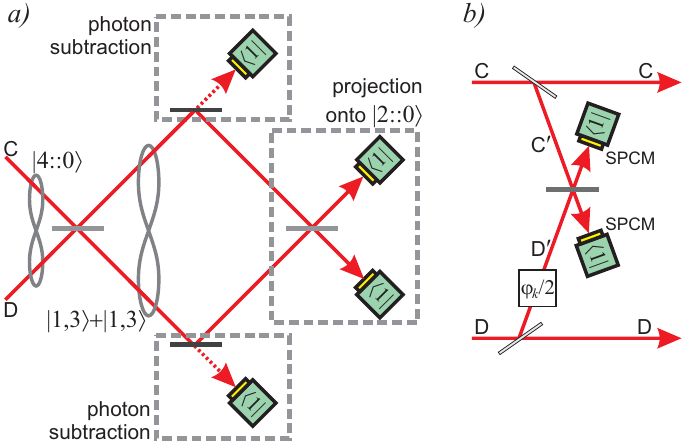}
	\caption{Projection onto N00N states with $N>2$. a) $N=4$. b) Implementation of operator $\hat c^2+e^{i\varphi_k}\hat d^2$ akin to the recipe of Ref.~\cite{Lee2001}.}
	\label{f1}
\end{figure}

A more general scheme, capable of detecting the N00N state with an arbitrary even $N$, involves applying operator $\hat c^N+\hat d^N$ to it. This will result in a two-mode vacuum state, which can be detected directly using SPCMs. Operator $\hat c^N+\hat d^N$ for even values of $N$ is realized using the scheme similar to that of Kok {\it et al.} \cite{Lee2001} using decomposition
\begin{equation}\label{cNdN}\hat c^N+\hat d^N=\prod_{k=1}^{N/2}(\hat c^2 +e^{i\varphi_k} \hat d^2),
\end{equation}where $\varphi_k=4\pi ik/N$. Operator $\hat c^2 +e^{i\varphi_kN} \hat d^2$ is implemented, in turn, via the setup shown in Fig.~1(b). The photons, tapped using weakly reflective beam splitters from both modes, are mixed on a symmetric beam splitter and detected using SPCMs. In this way, the state of the tapped modes C$'$ and $D'$ is projected onto the two-photon N00N state thanks to the reverse Hong-Ou-Mandel effect, so that one of the original modes loses two photons while the other one stays unaffected. A phase shift of $\varphi_k/2$ is applied to  mode D$'$ in order to account for factor $e^{i\varphi_k}$ in \eeqref{cNdN}.

In order to detect  state  $\ket{N::0}$, $N/2$ operators $\hat c^2 +e^{i\varphi_k} \hat d^2$ must be applied in sequence as per \eeqref{cNdN}. For low down-conversion amplitude, the probability for modes C and D to contain photons after this sequence is negligibly small, so the final double-vacuum detection is in fact not necessary. However, the probability for this scheme to successfully herald a N00N state decreases exponentially with $N$.


\begin{thebibliography}{99}

\bibitem{Our09}  Ourjoumtsev, A.,  Ferreyrol, F., Tualle-Brouri, R. \& Grangier P.
Preparation of non-local superpositions of quasi-classical light states.
{\it Nature Phys. } \textbf{5}, 189-192 (2009).

\bibitem{Dakic2012}
Daki\'c, B.  {\it et al.}
Quantum discord as resource for remote state preparation.
{\it Nature Phys.}  \textbf{8}, 666-670 (2012).

\bibitem{Lopaeva2013}
Lopaeva, E. D.  {\it et al.}
Experimental Realization of Quantum Illumination.
{\it Phys. Rev. Lett.} \textbf{110}, 153603 (2013).


\bibitem{Ulanov15} Ulanov, A. E. {\it et al.}
Undoing the effect of loss on quantum entanglement.
\textit{Nature Photon.}  \textbf{9}, 764-768 (2015).

\bibitem{Simon2003}
Simon, C. \& Irvine, W. T. M.
Robust Long-Distance Entanglement and a Loophole-Free Bell Test with Ions and Photons.
{\it Phys. Rev. Lett. } \textbf{91}, 110405 (2003).

\bibitem{Monroe} 
Moehring, D. L.  {\it et al.}
Entanglement of Single Atom Quantum Bits at a Distance. {\it Nature} \textbf{449}, 68-71 (2007).


\bibitem{Weinf} 
Hofmann, J. {\it et al.}
Heralded entanglement between widely separated atoms.
{\it Science} \textbf{337}, 72-75 (2012).

\bibitem{Hanson} 
Hensen, B. {\it et al.}.
Experimental loophole-free violation of a Bell inequality using entangled electron spins separated by 1.3 km.
{\it Nature} \textbf{526}, 682-686 (2015).

\bibitem{Lee2001} Kok, P., Lee, H. \& Dowling, J. P. Creation of large-photon-number path entanglement conditioned on photodetection. \textit{Phys. Rev. A} \textit{65}, 052104 (2002).

\bibitem{Edamatsu2002} Edamatsu, K., Shimizu, R. \& Itoh, T.
Measurement of the photonic de Broglie wavelength of entangled photon pairs generated by spontaneous parametric down-conversion.
{\it Phys. Rev. Lett.} \textbf{89}, 213601 (2002).



\bibitem{Nagata2007} Nagata, T., Okamoto, R., O'Brien, J. L., Sasaki, K. \& Takeuchi, S.
Beating the standard quantum limit with four-entangled photons.
{\it Science } \textbf{316}, 726-729 (2007).

\bibitem{Dowling2009} Dowling, J. P. Quantum Optical Metrology --- The Lowdown on High-N00N States.
{\it Contemporary Physics} \textbf{49}, 125-143 (2013).



\bibitem{Chen2015} Chen, P., Shu, C., Guo, X., Loy, M. M. T., \& Du, S.
Measuring the Biphoton Temporal Wave Function with Polarization-Dependent and Time-Resolved Two-Photon Interference.
{\it Phys. Rev. Lett.} \textbf{114}, 010401 (2015).


\bibitem{Kacprowicz2010}
Kacprowicz, M., Demkowicz-Dobrza\'{n}ski, R., Wasilewski, R., Banaszek, K. \& Walmsley, I. A.
Experimental quantum-enhanced estimation of a lossy phase shift.
{\it Nature Photon.} \textbf{4}, 357 - 360 (2010).

\bibitem{Rubin2007}
Rubin, M. A. \& Kaushik, S.
Loss-induced limits to phase measurement precision with maximally entangled states.
{\it Phys. Rev. A} \textbf{75}, 053805 (2007).

\bibitem{HOM87} Hong, C. K., Ou, Z. Y., \& Mandel, L.
Measurement of Subpicosecond Time Intervals between Two Photons by Interference.
{\it Phys. Rev. Lett.}  \textbf{59}, 2044-2046 (1987).



\bibitem{Ourjoumtsev2007}
Ourjoumtsev, A., Jeong, H., Tualle-Brouri,  R. \& Grangier P.
Generation of optical 'Schr\"odinger cats' from photon number states.
{\it Nature} \textbf{448}, 784-786 (2007).

\bibitem{AndersenRev} Andersen, U. L.,  Neergaard-Nielsen, J. S., van Loock, P. \& Furusawa, A. Hybrid discrete- and continuous-variable quantum information.
{\it Nature Phys.} \textbf{11}, 713–719 (2015).

\bibitem{Dowling2005} Dowling, J. P., Franson, J. D., Lee, H., Milburn, G. J. Towards scalable linear-optical quantum computers.
{\it Experimental Aspects of Quantum Computing}  \textbf{3}, 205-213 (2005)

\bibitem{Lim2005} Lim, Y. L., Beige, A., \& Kwek, L. C. Repeat-until-success linear optics distributed quantum computing.
{\it Phys. Rev. Lett.} \textbf{95}, 1-4 (2005).

\bibitem{Etesse2015} Etesse, J., Bouillard, M., Kanseri, B., \& Tualle-Brouri, R.
Experimental Generation of Squeezed Cat States with an Operation Allowing Iterative Growth.
{\it Phys. Rev. Lett.}  \textbf{114}, 1-5 (2015).

\bibitem{Walther04}  Walther, P. {\it et al.} De Broglie Wavelength of a Non-Local Four-Photon State. {\it Nature} \textbf{429}, 158--161 (2004).

\bibitem{Mitchell04} Mitchell, M. W., Lundeen, J. S. \& Steinberg, A. M. Super-Resolving Phase Measurements With a Multiphoton Entangled State. {\it Nature} \textbf{429},
161--164 (2004).






\bibitem{Pan1998} Pan, J.-W., Bouwmeester, D., Weinfurter, H. \&  Zeilinger, A. Experimental Entanglement Swapping: Entangling Photons That Never Interacted, {\it Phys. Rev. Lett.} \textbf{80}, 3891-3894 (1998).





\bibitem{SuppInfo} See Supplementary Information for details on extending our method to higher-photon number N00N states.

\bibitem{Berry2010}
Berry, D. W. \& Lvovsky, A. I.
Linear-Optical Processing Cannot Increase Photon Efficiency.
{\it Phys. Rev. Lett. }  \textbf{105}, 203601 (2010).



\bibitem{Lvovsky2009} Lvovsky,  A. I. \&  Raymer, M. G.
Continuous-variable optical quantum-state tomography.
{\it Rev. Mod. Phys. } \textbf{81}, 299 (2009).

\bibitem{Kumar2012}
Kumar, R. {\it et al.} Versatile wideband balanced detector for quantum optical homodyne tomography.
 {\it Opt. Commun.} \textbf{285}, 5259 (2012).

\bibitem{Lvovsky2001}
Lvovsky, A. I. {\it et al.}
Quantum State Reconstruction of the Single-Photon Fock State.
{\it Phys. Rev. Lett. } \textbf{87}, 050402 (2001).

\bibitem{Huisman2009}
Huisman, S. R.  {\it et al.}
Instant single-photon Fock state tomography.
{\it Optics Letters } \textbf{34}, 2739 - 2741 (2009).


\bibitem{Huang2015}
Huang, K. {\it et al.}
Optical synthesis of large-amplitude squeezed coherent-state superpositions with minimal resources.
{\it Phys. Rev. Lett.} \textbf{115}, 023602 (2015).

\bibitem{Lvo2004}
Lvovsky, A. I.
Iterative maximum-likelihood reconstruction in quantum homodyne tomography.
{\it J. Opt. B: Quantum Semiclass. Opt.}  \textbf{6}, S556-S559 (2004).

\bibitem{Lvo2007}
\v{R}eh\'{a}\v{c}ek, J., Hradil, Z., Knill, E., \& Lvovsky, A. I.
Diluted maximum-likelihood algorithm for quantum tomography.
{\it  Phys. Rev. A} \textbf{75}, 1-5 (2007).

\bibitem{Jeong14}
Jeong,	H. {\it et al.} Generation of hybrid entanglement of light.
{\it Nature Photon.} \textbf{8}, 564–569 (2014)

\bibitem{Morin14} Morin, O. {\it et al.}
Remote creation of hybrid entanglement between particle-like and wave-like optical qubits.
{\it Nature Photon.} \textbf{8}, 570–574 (2014).








\end{thebibliography}

\begin{thebibliography}{99}
\bibitem{Lee2001} Kok, P., Lee, H. \& Dowling, J. P. Creation of large-photon-number path entanglement conditioned on photodetection. Phys. Rev. A 65, 052104 (2002).
\end{thebibliography}
\end{document}